\documentclass{phb-proc4-auth}

\usepackage{graphicx}
\usepackage{amssymb}

\begin{document}
\begin{frontmatter}

%----------------------------------------------------------------------
% Specify destination and version number of the manuscript

\journal{SCES '04}

%----------------------------------------------------------------------
% Title of manuscript

\title{Point-contact investigations of challenging superconductors: two-band
MgB$_2$, antiferromagnetic HoNi$_2$B$_2$C, heavy fermion
UPd$_2$Al$_3$, paramagnetic MgCNi$_3$.}

\author{Yu. G. Naidyuk$^a$, O. E. Kvitnitskaya$^a$, I. K. Yanson$^a$,
G. Fuchs$^b$, K. Nenkov$^b$,}  \author{S.-L. Drechsler$^b$, G.
Behr$^b$, D. Souptel$^b$, and K. Gloos$^c$}

%----------------------------------------------------------------------
% List of authors
%
% List each author using a separate \author{} command
%
% If there is more than one author address, add a label to each author
% of the form \author[label]{name}.  This label should be identical to
% the corresponding label provided with the \address command.
%
% e.g. if there are three authors from two institutions in USA and
% France, you can link them to their respective addresses, using
%
% \author[US]{John Doe}
% \author[US,FR]{Jane Doe}
% \author[FR]{Jean Dupont}
% \address[US]{University of Life, Somewhere, USA}
% \address[FR]{Universite de la Vie, Quelque Part, France}
%
% N.B. Unlike the document class used for abstract submissions, it is
% possible to have the author associated with more than one address,
% as shown in the example above.

%\author[UA]{Yu. G. Naidyuk\corauthref{1}}
%\author[UA]{O. E. Kvitnitskaya}
%\author[UA]{I. K. Yanson}
%\author[GE]{G. Fuchs}
%\author[GE]{K. Nenkov}
%\author[GE]{S.-L. Drechsler}
%\author[GE]{G. Behr}
%\author[GE]{D. Souptel}
%\author[DE]{K. Gloos}

%----------------------------------------------------------------------
% If there is more than one address, list each using a separate
% \address command using a label to link it to the respective author
% as described above

\address[UA]{B. Verkin Institute for Low Temperature Physics
and Engineering, 47 Lenin Ave., 61103,  Kharkiv, Ukraine}
\address[GE]{Leibniz-Institut f\"ur Festk\"orper- und
Werkstofforschung Dresden e.V., Postfach 270116, D-01171 Dresden,
Germany}
\address[DE]{Nano-Science Center, Niels Bohr Institute fAFG,
Universitetsparken 5, DK-2100 Copenhagen, Denmark}
%----------------------------------------------------------------------
% Title page footnotes
%
% If you need to add qualifying information to any of the authors,
% use the \thanksref{} command within the \author command.  The
% argument is the label of a corresponding \thanks[label]{text}
% command which contains the footnote text
%
% e.g. you can acknowledge a funding authority for John Doe, using
%
% \author{John Doe\thanksref{ABC}}
% \thanks[ABC]{This work was supported by Institute of Unphysical
%    Phenomena under contract no. ABC-123}
%

%\thanks[]{}

%----------------------------------------------------------------------
% Contact Information
%
% Add the complete postal address, telephone number, fax number, and
% email address of the corresponding author as a special footnote using
% the \corauth[]{} command.  This works in a similar way to the \thanks
% command.  Add the \corauthref{} command within the \author command.
% The argument is the label of a corresponding \corauth[label]{text}
% command which contains the contact information.  Prefix the text with
% Corresponding Author:
%
% e.g. if the contact author is John Doe,
%
% \author{John Doe\corauthref{1}}
% \corauth[1]{Corresponding Author: University of Life, 123 Some St.,
%    Somewhere, MI 12345, USA.  Phone: (555) 555-5555
%    Fax: (555) 555-7777, Email: JDoe@uol.edu}
%

\corauth[1]{Corresponding Author:Email:naidyuk@ilt.kharkov.ua}

%----------------------------------------------------------------------

\begin{abstract}

An overview on recent efforts in point-contact (PC) spectroscopy
of title superconductors is given. Distinct phonon features and
crystalline-electric-field effects are observed in PC spectra of
HoNi$_2$B$_2$C. Results of study of superconducting (SC) gap and
excess current versus temperature and magnetic field reflecting
specific multi-band electronic structure in MgB$_2$ are presented.
The nature of the extremely nonlinear $I(V)$ curves in the
antiferromagnetic (AF) and SC state are elucidated for
UPd$_2$Al$_3$ break-junctions and MgCNi$_3$ point contacts.

\end{abstract}

%----------------------------------------------------------------------
% Manuscript keywords
%
% Please give two or three keywords in the form: keyword \sep keyword
% e.g. NMR \sep superconductivity
%
% NB The syntax is different from the abstract document class


%----------------------------------------------------------------------
% End of front page

\end{frontmatter}

By point-contact (PC) investigations both the superconducting (SC)
order parameter and PC electron-phonon interaction (EPI) function
$\alpha^2_{\rm PC}2F(\omega)$ can be established studying the
first and second derivatives of the $I(V)$ characteristic of PC's
\cite{Naidyuk}. Thus the PC spectroscopy could be helpful to
illuminate details of EPI and characteristic of SC state in the
title compounds.

%{\it Experimental details}. We have used single crystals of grown
%by the rf-heated floating-zone method. The residual resistivity
%was $\rho_0=0.33\mu\Omega$\,cm and residual resistivity ratio 24.
%Further details of the PC measurements are given elsewhere
%\cite{Naidyuk02}.

\begin{figure}
\begin{center}
\includegraphics[width=8cm,angle=0]{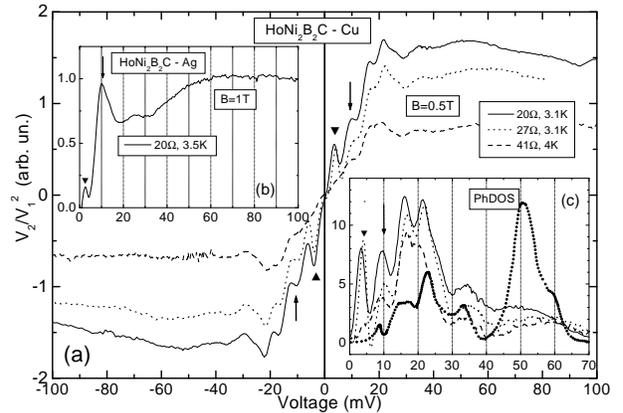}
\end{center}
\caption[] {PC spectra ($V_2/V_1^2\propto -d^2I/dV^2(V)
\propto\alpha_{PC}^2 (\epsilon)\,F(\epsilon)$) of several
HoNi$_2$B$_2$C--Me (Me=Cu,\,Ag) PCs. Magnetic field is applied to
suppress superconductivity. Insets: (b) spectrum with main 10-mV
CEF peak. $\downarrow$ marks CEF peak, $\blacktriangledown$ marks
``mag\-netic'' peak about 4 mV, (c) spectra from the main panel
with  subtracted background along with LuNi$_2$B$_2$C PhDOS
\cite{Gompf}.} \label{honibc}
\end{figure}

We have measured PC spectra of HoNi$_2$B$_2$C with pronounced
phonon maxima at about 16 and 22\,mV, a smeared maximum near
34\,mV, and shoulder around 50\,mV (Fig.\,\ref{honibc}). All these
features correspond well to the neutron phonon DOS of nonmagnetic
related compound LuNi$_2$B$_2$C \cite{Gompf}, only the high energy
part of the PC spectrum is remarkably smeared. The maximum around
10\,mV might be connected with CEF excitations, observed in this
range by neutron scattering \cite{Gasser}, while the maximum
around 4\,mV has definitely connection with magnetic order, since
it disappears above the Ne$\acute{e}$l ($\sim$6K) temperature.
More often the spectra demonstrate completely smeared phonon
maxima, but expressed CEF peak (see Fig.\,\ref{honibc}b). This
points to the importance of CEF excitations in the transport as
well as in the SC properties of HoNi$_2$B$_2$C.
\begin{figure}
\begin{center}
\includegraphics[width=7cm,angle=0]{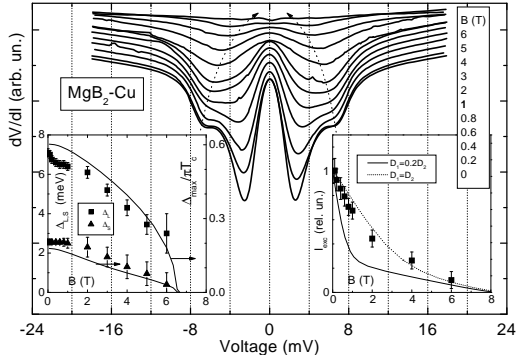}
\end{center}
\caption[] {$dV/dI$ of a MgB$_2$--Cu contact at 4.5K in a magnetic
field. Dashed arrows follow qualitatively the large gap evolution.
Large and small gap (left inset) and excess current (proportional
to the integral intensity of $dV/dI$ minima, right inset) vs
magnetic field for the contact from the main panel along with
theoretical prediction (solid curves) calculated in
\cite{NaidyukF}.} \label{mgb2}
\end{figure}

MgB$_2$ has multi-band electronic structure with SC gaps
distributed over the Fermi surface being $\Delta_\sigma\approx$
7\,meV for the $\sigma$-band and $\Delta_\pi\approx$ 2\,meV for
the $\pi$-band \cite{YansonR}. This is seen in the $dV/dI$ curves
exhibiting two sets of minima (Fig.\,\ref{mgb2}). We have measured
magnetic field dependence of both gaps along with an excess
current. The latter has positive curvature, unlike for common
superconductors, which is connected  \cite{NaidyukF} with the
specific multi-band electronic structure of MgB$_2$.

\begin{figure}
\begin{center}
\includegraphics[width=7cm,angle=0]{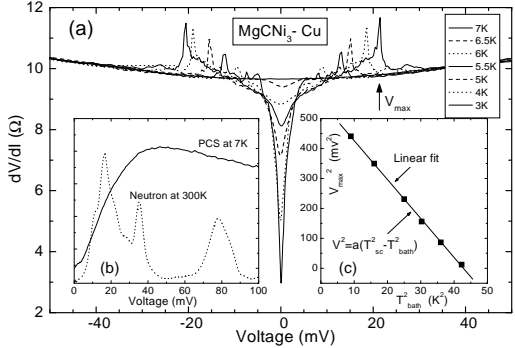}
\end{center}
\caption[] {$dV/dI$ of a MgCNi$_3$-Cu heterocontact below T$_c
\simeq$7\,K. Insets: (b) second derivative of $I(V)$ curve for the
same contact just above T$_c$ in comparison with smoothed phonon
DOS \cite{Heid}, (c) peak position in $dV/dI$ vs temperature.}
\label{mgcni}
\end{figure}

MgCNi$_3$ becomes superconducting below about 8\,K despite the
high content of magnetic Ni, which may favor an unconventional
pairing mechanism. The large residual resistivity $\rho_0$ of
MgCNi$_3$, like that in amorphous metal, requires at first an
ascertainment of the current flow regime in PC. As it is discussed
in \cite{NaidyukC} some distinct features as spikes and zero-bias
minimum in $dV/dI$ of MgCNi$_3$ PCs in SC state
(Fig.\,\ref{mgcni}) have relation neither to order parameter (or
gap) nor to unconventional ground state. The regime of current
flow in MgCNi$_3$ PCs is likely thermal and PC spectrum shows no
discernible phonon features (Fig.\,\ref{mgcni}b).
\begin{figure}
\begin{center}
\includegraphics[width=6.5cm,angle=0]{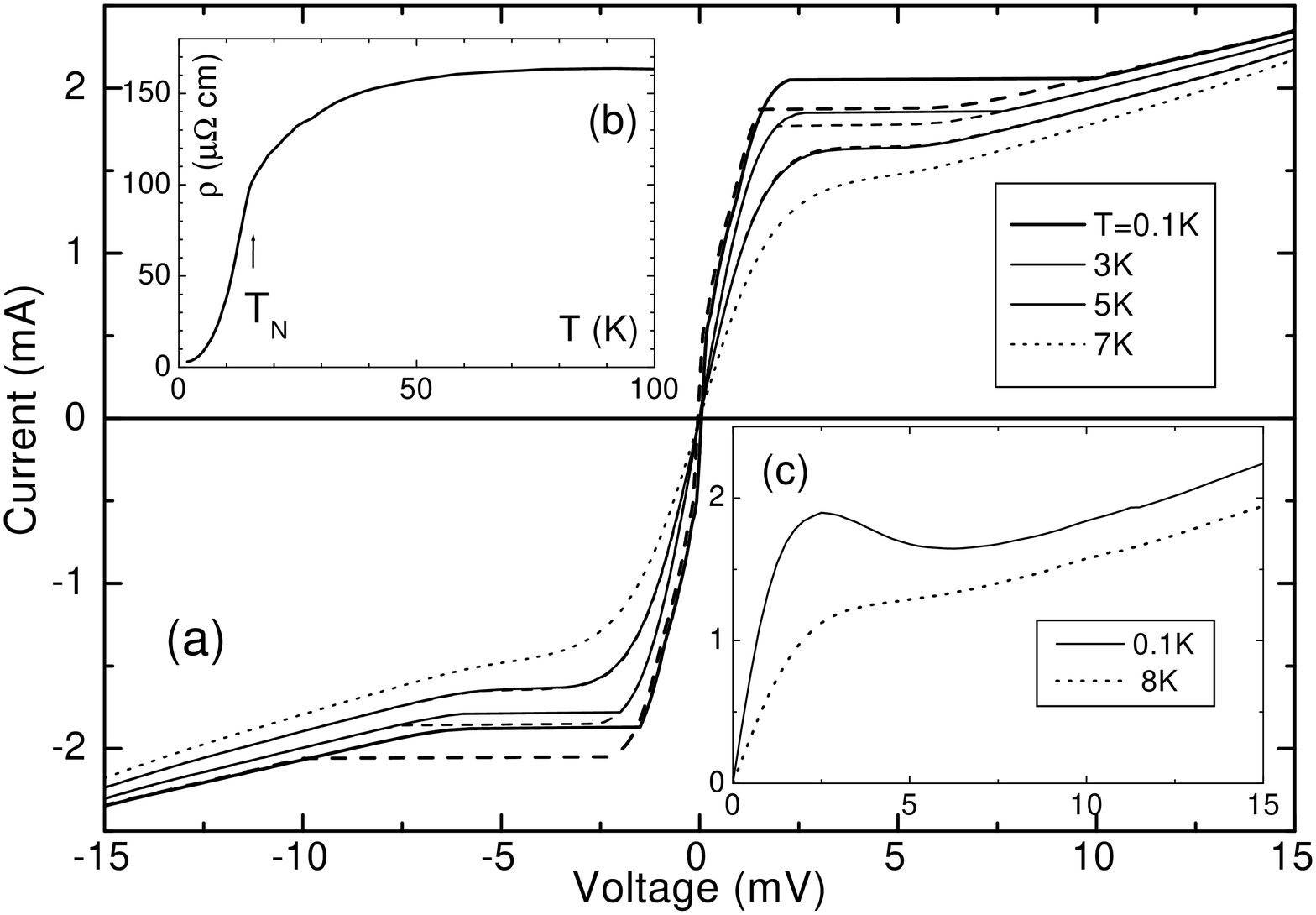}
\end{center}
\caption[] {$I(V)$ characteristics of a UPd$_2$Al$_3$ break
junction with $R_n = 0.66\,\Omega$ at the indicated temperatures.
Solid (dashed) lines correspond to sweeps with increasing
(decreasing) current. The hysteretic loops become smaller when the
temperature rises and vanish above $\sim 5\,$K. Inset: (b)
$\rho(T)$ for the bulk compound, (c) $I(V)$ characteristics of the
UPd$_2$Al$_3$ PC at two temperatures, calculated for the thermal
regime \cite{NaidyukU}} \label{updal}
\end{figure}

UPd$_2$Al$_3$ has much lower  $\rho_0$ compared to MgCNi$_3$,
however $\rho(T)$ increases steeply by approaching the AF
transition at 14\,K (Fig.\,\ref{updal}b). We have observed that
sub-micron PCs of UPd$_2$Al$_3$ have {\em hysteretic} $I(V)$
characteristics when the junction is driven by a current source
(Fig.\,\ref{updal}). It turned out that $I(V)$ can be reproduced
theoretically by assuming the constriction to be in the thermal
regime \cite{NaidyukU}. Thus, these PCs represent non-linear
devices with $N$-shaped $I(V)$ curves (Fig.\,\ref{updal}c) that
have a negative differential resistance. Such behavior can be
expected for PCs with other magnetic materials, which resistivity
increases steeply when magnetic order is destroyed by thermal
fluctuations.

The DFG (SFB 463) support and US CRDF (No.UP1-2566-KH-03) grant
are acknowledged.
%----------------------------------------------------------------------
% Reference section
% List each reference with a separate \bibitem{} command.  The
% argument contains the label that is used in the \cite{} command
% in the main text
% e.g.  This follows our pioneering work on TdB2\cite{TdB2}.
% \bibitem{TdB2}
% J. Doe, J. Doe, and J. Dupont, J. Irrep. Res. 10 (2000) 1000.

%----------------------------------------------------------------------
% Figures and Tables
% Insert figures and tables at the end of the document unless you
% are familiar with the LaTeX positional options.
% \begin{figure}
%     \centering
%     \includegraphics{filename.eps}
%     \caption{Insert figure caption here}
% \end{figure}
%
% \begin{table}
%     \centering
%     \begin{tabular}
%     Insert table here
%     \end{tabular}
%     \caption{Insert table caption here}
% \end{table}
% Please refer to other LaTeX documentation for help on using these
% environments.
%----------------------------------------------------------------------
% Terminate document

\begin{thebibliography}{00}

%\bibitem{Yanson} I. K. Yanson, Sov. J. Low Temp. Phys. {\bf 9} (1983) 343.

\bibitem{Naidyuk} Yu. G. Naidyuk, I.~K.~Yanson,  {\it Point-contact spectroscopy}
(Springer, 2004).

\bibitem{Gompf} F. Gompf et al., Phys. Rev. B {\bf 55} (1997) 9058.

\bibitem{Gasser}U. Gasser et al., Z. Phys. B {\bf 101} (1996) 345;
N. Cavadini et al., Eur. Phys. J B {\bf 29} (2002) 377.

\bibitem{Heid} R. Heid et al., Phys. Rev. B {\bf 69} (2004) 092511.

\bibitem{YansonR} I. K. Yanson et al., Low Temp. Phys. {\bf 69} (2004) 261.

\bibitem{NaidyukF}Yu. G. Naidyuk et al., cond-mat/0403324.

%\bibitem{Verkin} B. I. Verkin et al., Solid State Commun. {\bf 30} (1979) 215.
%Izv.~Akad.~Nauk SSSR, Ser.~Fiz. {\bf 44}, 1330 (1980).

%\bibitem{Kulik} I.~O.~Kulik, Sov.~J.~Low Temp.~Phys. {\bf 18} (1992) 302.

\bibitem{NaidyukU} Yu. G. Naidyuk et al., J.Phys.:Cond. Mat. {\bf 16} (2004)
3433.

\bibitem{NaidyukC}Yu. G. Naidyuk, Phys. Rev. B  {\bf 69} (2004) 136501.

%\bibitem{vanGelder80}  A. P. van Gelder, Solid State Commun. {\bf 35} 19  (1980).

%\bibitem{Zr} PC spectra of Zr were put at our disposal by N. L. Bobrov and V. V. Fisun
%(unpublished).

\end{thebibliography}
\end{document}